\documentstyle[12pt,psfig,amssymb,amsmath]{article}

\begin{document}

\title{Rayleigh Instability in Liquid Crystalline Jet}

\author{Leonid G. Fel and Yoram Zimmels\\
\\Department of Civil Engineering, Technion, \\Haifa 32000, Israel}
\maketitle

\def\be{\begin{equation}}
\def\ee{\end{equation}}
\def\p{\prime}

\begin{abstract}
The capillary instability of liquid crystalline (LC) jets is considered 
in the framework of linear hydrodynamics of uniaxial nematic LC. The free
boundary conditions of the problem are formulated in terms of mean surface 
curvature ${\cal H}$ and Gaussian surface curvature ${\cal G}$. The static 
version of capillary instability is shown to depend on the elasticity modulus 
$K$, surface tension $\sigma_0$, and radius $r_0$ of the LC jet, 
as expressed by the characteristic parameter $\varkappa=K/\sigma_0 r_0$.
The problem of capillary instability in LC jets is solved exactly and 
a dispersion relation, which reflects the effect of elasticity, is derived. 
It is shown that increase of the elasticity modulus results in a decrease 
of both the cut off wavenumber $k$ and the disturbance growth rate $s$. 
This implies enhanced stability of LC jets, compared to ordinary liquids.
In the specific case, where the hydrodynamic and orientational LC modes 
can be decoupled, the dispersion equation is given in closed form.
\end{abstract}  

\vskip .5cm
\centerline{{\bf Key words:} Jet, Rayleigh Instability, Nematic Liquid Crystal}

\vskip 2.5cm
\centerline{e-mail: lfel@techunix.technion.ac.il}
\newpage

\noindent
\noindent
\section{Introduction}
\label{intro}
The breakup of liquid jets, that are injected through a circular nozzle into stagnant
fluids, has been the subject of widespread research over the years. Previous studies 
that followed the seminal works of Lord Rayleigh have established that the complex jet 
flow is influenced by a large number of parameters. These include nozzle internal flow
effects, the jet velocity profile and the physical state of both liquid and gas. 
Notwithstanding the fact that the hydrodynamic equations are nonlinear, the linear stability 
theory can provide qualitative descriptions of breakup phenomena and predict the existence 
of different breakup regimes.

Rayleigh showed \cite{Rayleigh79} by using a linear theory that the jet breakup is a 
consequence of hydrodynamic instability, or more exactly {\it capillary instability}. 
Neglecting the effect of the ambient fluid, the viscosity of the jet liquid, and 
gravity, he demonstrated that a cylindrical liquid jet is unstable with respect to 
disturbances characterized by wavelengths larger than the jet circumference. He also 
considered the case of a viscous jet in an inviscid gas and an inviscid gas jet in an 
inviscid liquid \cite{Rayleigh92}. Weber \cite{Weber31} generalized Rayleigh's result for 
the case of a {\it Newtonian} viscous liquid and showed that the viscosity tends to 
reduce the breakage rate and increase the drop size. Chandrasekhchar \cite{Chandra61} 
considered the effect of a uniform magnetic field on the capillary instability of a 
liquid jet. A mechanism of bending disturbances and of buckling, slowly moving, highly 
viscous jets, was presented by Taylor \cite{Taylor69}. Further developments of the theory 
in Newtonian liquids was concerned with additional factors such as the dynamic action of the 
ambient gas (leading to atomization of the jet), the nonlinear interaction of growing 
modes that lead to satellite drop formation, and the spatial character of instability 
(see \cite{Levich62}, \cite{Yarin93}). 

The capillary instability in jets, comprised of {\it non--Newtonian} suspensions and 
emulsions, presents a different category of cases which are governed by 
power--law 
(pseudoplastic and dilatant) liquids.  The effective viscosity of the pseudoplastic 
liquid decreases with growth of strain rate, whereas in dilatant liquids, 
it increases \cite{Yarin93}. The behaviour of capillary jets of dilute and concentrated 
polymer solutions suggest a strong influence of the macromolecular coils on their  
flow patterns \cite{Yarin93}.  Free jets of polymeric liquids, that exhibit oscillations, 
are reported in \cite{Tomito79}. 

Recently the idea of Rayleigh instability was applied to tubular membranes in dilute 
lyotropic phases \cite{Moses94}. Their relaxation, following optical excitation, 
is characterized by a long time, and can be described by means of hydrodynamic approach 
\cite{lebed96}. Bending deformations of such membranes are governed by the Helfrich 
energy \cite{helfr75} which depends on the curvature of the tube. Thus, competition
between the surface tension and curvature energy of the water immersed membrane 
renders the initial shape of the tube unstable. The hydrodynamic formalism used 
in \cite{lebed96} and the hydrodynamics of fluids with inner order such as liquid 
crystal (LC) \cite{Kats94} have similar features. In \cite{lebed96} the order 
parameter stands for a unit vector normal to the membrane surface. In contrast, the 
order parameter $Q$ of a LC fluid, is defined throughout the space it occupies.

The continuum theory of LC phases has emerged as a rigorous part of condensed 
matter theory. The hydrodynamics of the LC phases was developed during the 
70--80th and its predictions were successfully confirmed in many experimental 
observations. The combination of viscous and elastic properties is likely to
produce new evolution patterns of hydrodynamic instabilities, in the context of
Benard--Rayleigh, Marangoni and electrohydrodynamic effects \cite{Genn93}, 
which cannot occur in ordinary liquids. However, its capillary instability, 
when in the form of a jet, was not considered as yet. In particular we refer 
to the {\it uniaxial nematic} phase. 
 
The instability of a LC jet poses an additional challenge with respect to the effects 
listed above. This applies already within the framework of linear stability theory. 
The LC class of fluids seems to provide a good example of unique properties, as 
compared to polymer solutions. The elastic properties of a LC are expected to change 
the evolution patterns of jets which are made from them. In this work we 
derive a rigorous mathematical 
model of capillary instability for isothermal incompressible nematic LC jets. 
This model shows how the combined viscous and elastic properties of LC fluids 
determine the boundary conditions at the free surface, and the range where 
instability prevails.

\section{Hydrodynamics of a liquid crystalline jet}
\label{hydro}
In this Section, we formulate first the problem of capillary instability and then 
derive the basic equations which govern the linear hydrodynamics of a liquid crystalline 
jet. The flow of a nematic LC is described by a set of differential equations 
supplemented by boundary conditions on the LC free surface: continuity equation, 
Navier--Stokes equation of visco--elastic LC, and Lesli--Ericksen equation of 
angular motion of the director ${\bf n}({\bf r},t)$.

The basic notations and linear hydrodynamic equations of uniaxial nematic liquid 
crystals follow the theory given in \cite{Kats94}, \cite{Luben71}, \cite{Lif86}.
\subsection{Basic notations and variables}
\label{variabl}
The following basic variables describe the nematic LC medium: 
velocity ${\bf V}({\bf r},t)$, pressure $P({\bf r},t)$ and LC--director 
${\bf n}({\bf r},t)$. The initial values of the functions will be denoted  
by "o", either as a subscript or superscript. The following notations, which are 
commonly accepted in the theory of LCs, are used henceforth:

1. The free energy density $E_d$ of deformed non-chiral uniaxial nematic LC, 
given in quadratic approximation in terms of the derivatives 
$\partial {\bf n}/\partial x_j$ reads 
\begin{eqnarray}
2E_d= K_1\;{\rm div}^{2}\;{\bf n} + K_2\langle{\bf n},{\rm rot}\;{\bf n}\rangle^{2} + 
       K_3 \left[{\bf n}\times {\rm rot}\;{\bf n}\right]^2\;,
\label{frank1}
\end{eqnarray}
where $\langle {\bf a},{\bf b}\rangle$ and $[{\bf a}\times {\bf b}]$ denote 
scalar and vector products of vectors, and $K_i\geq 0,i=1,2,3$ are known 
as the Frank elasticity moduli. In the vicinity of a phase transition 
$K_i\propto Q^2$ \cite{Genn93} and in the isotropic phase they vanish.

2. The bulk molecular field ${\bf F}$ and the Ericksen elastic stress 
tensor $\tau_{ki}$, which set the equilibrium distribution of the 
${\bf n}$--field in a LC, are determined by the following variational derivatives
\footnote{Here and throughout the paper, unless noted otherwise, we 
apply the summation rule over indices which are repeated in a tensor  
product, e.g. $a_{ij} b_{jk}=\sum_j a_{ij} b_{jk}$.}
\begin{eqnarray}
{\bf F}={\bf M}-{\bf n}\langle{\bf n},{\bf M}\rangle\;,\;\;\;
\mbox{or}\;\;\;F_i=(\delta_{ij}-n_i n_j) M_j\;,
\label{frank1a}
\end{eqnarray}
where
\begin{eqnarray}
M_i&=&\frac{\partial }{\partial x_k}\;\frac{\partial E_d}
{\partial (\partial_k n_i)}-\frac{\partial E_d}{\partial n_i}
\;\;,\;\;\;
\tau_{ki}=\frac{\partial E_d}{\partial (\partial_k n_i)}\;,\;\;\;
\partial_k =\frac{\partial }{\partial x_k}\;,
\label{frank2}
\end{eqnarray}
i.e. 
\begin{eqnarray}
{\bf M}&=&K_1\;{\rm grad}\;{\rm div}\;{\bf n}-
K_2 \left\{\langle{\bf n},{\rm rot}\;{\bf n}\rangle {\rm rot}\;{\bf n}+
{\rm rot}\left(\langle{\bf n},{\rm rot}\;{\bf n}\rangle {\bf n}\right)\right\}+
\nonumber\\
&&K_3 \{{\rm rot}\;\left[{\bf n}\times \left[{\bf n}\times {\rm rot}\;
{\bf n}\right]\right]+
\left[\left[{\bf n}\times {\rm rot}\;{\bf n}\right]\times {\rm rot}\;{\bf n}
\right]\}\;,\nonumber\\ 
\tau_{ki}&=&K_1\;\delta_{ki}{\rm div}\;{\bf n} +
K_2 \langle{\bf n},{\rm rot}\;{\bf n}\rangle n_m \epsilon_{mki}+
K_3 \left[\left[{\bf n}\times {\rm rot}\;
{\bf n}\right]\times {\bf n}\right]_m \epsilon_{mki}\;,
\label{frank3}
\end{eqnarray}
$\epsilon_{mki}$ is a completely antisymmetric unit tensor of the 3rd rank 
(Levi--Civita tensor). 

3. If the deviations of the director ${\bf n}={\bf n}^0+{\bf n}^1$ 
from its initial orientation ${\bf n}^0$ are small, then  
\begin{eqnarray}
n^0_x=n^0_y=0,\;n^0_z=1\;,\;\;
1\gg n^1_x, n^1_y\gg n^1_z \sim \left(n^1_x\right)^2
,\left(n^1_y\right)^2\;,
\label{frank3d}   
\end{eqnarray}
and simple algebra yields the following linear approximation
\begin{eqnarray}
M_x={\widehat {\sf K}}n^1_x+\left(K_1-K_2\right)
\frac{\partial^2 n^1_y}{\partial x \partial y}\;,\;
M_y={\widehat {\sf K}}n^1_y+\left(K_1-K_2\right)
\frac{\partial^2 n^1_x}{\partial x \partial y}\;,\;
M_z=\left(K_1-K_3\right)\frac{\partial }{\partial z}
{\rm div}\;{\bf n}^1\;,
\label{frank3a}
\end{eqnarray}
where ${\widehat {\sf K}}=K_1\frac{\partial^2}{\partial x^2}+
K_2\frac{\partial^2}{\partial y^2}+K_3\frac{\partial^2}{\partial z^2}$, 
and by virtue of (\ref{frank1a}), $F_x=M_x\;,\;\;F_y=M_y\;,\;\;F_z=0$. If 
further simplification through single elastic approximation $K_1=K_2=K_3=K$ is 
applied, then 
\begin{eqnarray}
F_x=K\Delta_3 n^1_x\;,\;\;F_y=K\Delta_3 n^1_y\;,\;\;F_z=0\;,
\;\;\;\;\Delta_3=\frac{\partial^2}{\partial x^2}+
\frac{\partial^2}{\partial y^2}+\frac{\partial^2}{\partial z^2}\;,
\label{frank3b}
\end{eqnarray}
where $\Delta_3$ is the three--dimensional Laplacian. Similar considerations 
regarding the Ericksen stress tensor $\tau_{ki}$ give
\begin{eqnarray}
\tau_{xx}=\tau_{yy}=\tau_{zz}=
K_1\;{\rm div}\;{\bf n}^1\;,\;\;
\tau_{xy}=-\tau_{yx}=K_2 \left(\frac{\partial n_y^1}{\partial x}-
\frac{\partial n_x^1}{\partial y}\right)\;,\nonumber\\
\tau_{yz}=-\tau_{zy}=K_3\left(\frac{\partial n_z^1}{\partial y}-
\frac{\partial n_y^1}{\partial z}\right)\;,\;\;
\tau_{zx}=-\tau_{xz}=K_3\left(\frac{\partial n_x^1}{\partial z}-
\frac{\partial n_z^1}{\partial x}\right)\;,
\label{frank3c}
\end{eqnarray}
The stresses given by (\ref{frank3c}) do not contribute to the non--dissipative 
stress tensor $T_{ik}^{\sf r}$ used in the linear hydrodynamics of LCs (see 
(\ref{frank4}) below).

4. The reactive (non--dissipative) $T_{ik}^{\sf r}$ and 
dissipative $T_{ik}^{\sf d}$ stress tensors are defined as follows
\begin{eqnarray}
T_{ik}^{\sf r}&=&-P\;\delta_{ik}
-\tau_{kj}\frac{\partial n_j}{\partial x_i}-
\frac{\lambda}{2}(n_i F_k+n_k F_i)+
\frac{1}{2}(n_i F_k-n_k F_i)\;,\label{frank4}\\
T_{ik}^{\sf d}&=&2\eta_1\Upsilon_{ik}+(\eta_2-\eta_1)\;\delta_{ik}
{\rm div}{\bf V}+(\eta_1-\eta_2+\eta_4)(\delta_{ik} n_j\Upsilon_{jm}n_m+
n_i n_k{\rm div}{\bf V})+\label{frank4a}\\
&&(\eta_3-2\eta_1) \left(n_i \Upsilon_{kj}n_j+n_k \Upsilon_{ij}n_j\right)+
(\eta_1+\eta_2+\eta_5-2\eta_3-2\eta_4)n_i n_k 
n_jn_m\Upsilon_{jm}\;,\nonumber
\end{eqnarray}
where the antisymmetric $\Omega_{ik}$ ({\it vorticity}) and symmetric 
$\Upsilon_{ik}$ parts of the derivative $\partial_k V_i$ read
\begin{eqnarray}
\Omega_{ik}=\frac{1}{2}\left(\frac{\partial V_k}{\partial x_i}-
\frac{\partial V_i}{\partial x_k}\right)\;,\;\;\;\;
\Upsilon_{ik}=\frac{1}{2}\left(\frac{\partial V_k}{\partial x_i}+
\frac{\partial V_i}{\partial x_k}\right)\;,
\label{frank5}
\end{eqnarray}
Five independent viscous moduli $\eta_j$, kinetic coefficient $\lambda$, and 
rotational viscosity $\gamma_1$, determine the dissipative stress tensor 
$T_{ik}^{\sf d}$, the 4th--rank viscosity tensor $\eta_{ikjm}$, and 
the dissipative function $D$ in the absence of heat fluxes
\begin{eqnarray}
D&=&\eta_{ikjm}\Upsilon_{ik}\Upsilon_{jm}+\frac{1}{\gamma_1}{\bf F}^2\;,
\;\;\;\;T_{ik}^{\sf d}=\eta_{ikjm}\Upsilon_{jm}\;,\label{frank6}\\
\eta_{ikjm}&=&\eta_1(\xi_{ij}\xi_{km}+\xi_{kj}\xi_{im})+(\eta_2-
\eta_1)\xi_{ik}\xi_{jm}+\frac{\eta_3}{2} (n_in_j\xi_{km}+n_kn_j\xi_{im}+
n_in_m\xi_{kj}+\nonumber\\
&&+n_kn_m\xi_{ij})+\eta_4 (n_in_k\xi_{jm}+n_jn_m\xi_{ik})+\eta_5n_in_kn_jn_m\;,
\;\;\;\xi_{ik}=\delta_{ik}-n_in_k\;.\nonumber
\end{eqnarray}
The tensor $\eta_{ikjm}$ consists of five independent uniaxial invariants 
\cite{Kats94} and is highly symmetrical $\eta_{ikjm}=\eta_{kimj}=\eta_{jmik}$. 
The requirement that $D$ be positive translates into, 
\begin{eqnarray}
\eta_1\geq 0\;,\;\;\eta_2\geq 0\;,\;\;\eta_3\geq 0\;,\;\;\eta_5\geq 
0\;,\;\;\eta_2\eta_5\geq \eta_4^2\;,\;\;\gamma_1\geq 0\;.
\label{lesli1}
\end{eqnarray}
The parameter $\lambda$ is close to $+1$ or $-1$ for rod--like or disk--like 
molecules, respectively. If the liquid is visco--isotropic, then $\lambda=0$.

5. The hydrodynamic reactive (non--dissipative) ${\bf m}^{\sf r}$ and 
dissipative ${\bf m}^{\sf d}$ fields are defined as follow
\begin{eqnarray}
m_i^{\sf r}=-\langle {\bf V}, \nabla_3\rangle n_i+
n_k\Omega_{ki}+\lambda\; \xi_{ij}\Upsilon_{jk}n_k\;,\;\;\;
{\bf m}^{\sf d}=\frac{1}{\gamma_1} {\bf F}\;,
\label{frank8}
\end{eqnarray}
where $\nabla_3$ is the three-dimensional gradient operator, 
$(\nabla_3)^2=\Delta_3$.

6. The surface tension $\sigma$ of nematic LC is given by \cite{Rapin71}, 
\begin{eqnarray}
\sigma=\sigma_0+\sigma_1\langle {\bf n},{\bf e}\rangle^2 \;,
\label{frank9}  
\end{eqnarray}
where $\sigma_0$ and $\sigma_1$ are isotropic and anisotropic surface tension moduli 
respectively, and ${\bf e}$ is a unit normal vector to the LC surface.

7. In the case of an incompressible LC 
$(\eta_1^{\sf in}=\eta_2^{\sf in},\;\eta_4^{\sf in}=0)$ the tensors $T_{ik}^{\sf 
d}$ and $\eta_{ikjm}$ take the following form,  
\begin{eqnarray}
T_{ik}^{\sf din}&=&2\eta_1^{\sf in}\Upsilon_{ik}+
(\eta_3^{\sf in}-2\eta_1^{\sf in}) \left(n_i \Upsilon_{kj}n_j+
n_k \Upsilon_{ij}n_j\right)+(2\eta_1^{\sf in}+\eta_5^{\sf in}-
2\eta_3^{\sf in})n_i n_k n_jn_m\Upsilon_{jm}\;\label{frank19}\\
\eta_{ikjm}^{\sf in}&=&\eta_1^{\sf in}(\xi_{ij}\xi_{km}+\xi_{kj}\xi_{im})+
\frac{\eta_3^{\sf in}}{2} (n_in_j\xi_{km}+n_kn_j\xi_{im}+n_in_m\xi_{kj}+
n_kn_m\xi_{ij})+\eta_5^{\sf in}n_in_kn_jn_m\;,\nonumber
\end{eqnarray}
where "${\sf in}$" denotes incompressibility condition.

8. The first two terms in (\ref{frank4a}) and the second equation of (\ref{frank6}) 
correspond to ordinary compressible liquids with isotropic invariance. 
The simplification of (\ref{frank4a}) results from
$$
\eta_3^{\sf L}=2 \eta_1^{\sf L}\;,\;\;
\eta_4^{\sf L}=\eta_2^{\sf L}-\eta_1^{\sf L}\;,\;\;
\eta_5^{\sf L}=\eta_2^{\sf L}+\eta_1^{\sf L}\;, 
$$
so that
\begin{eqnarray}
T_{ik}^{\sf dL}=2\eta_1^{\sf L}\Upsilon_{ik}+
(\eta_2^{\sf L}-\eta_1^{\sf L})\;\delta_{ik}{\rm div}{\bf V}\;,\;\;\;
\eta_{ikjm}^{\sf L}=\eta_1^{\sf L}(\delta_{ij}\delta_{km}+\delta_{im}\delta_{kj})+
(\eta_2^{\sf L}-\eta_1^{\sf L})\delta_{ik}\delta_{jm}\;.\nonumber
\end{eqnarray}
The coefficients $\eta_1^{\sf L}$ and $\eta_2^{\sf L}-\eta_1^{\sf L}$ are known 
as {\it the first} and {\it second} isotropic viscosities.

9. Another system of viscous moduli $\alpha_i$ (called {\it Lesli 
viscosities}) relate dissipative and kinetic moduli in the following way
\footnote{The correct expression for $\eta_5$ is given in \cite{Luben71}.}
\begin{eqnarray}
&&\eta_1=\frac{\alpha_4}{2}\;,\;\;\;
\lambda=-\frac{\gamma_2}{\gamma_1}\;,\;\;\;
\eta_5=\alpha_1+\alpha_4+\alpha_5+\alpha_6\;,\;\;\;
\gamma_1=\alpha_3 -\alpha_2,\;\;\;\gamma_2=\alpha_3 +\alpha_2\;,\nonumber\\
&&\eta_3-2\eta_1=\alpha_5+\alpha_2\lambda\;,\;\;\;
2\eta_1+\eta_5-2\eta_3=\alpha_1 +\frac{\gamma_2^2}{\gamma_1}\;,
\label{lesli2}
\end{eqnarray}
with the support of Onzager--Parodi relation \cite{Par75} 
$\alpha_3 +\alpha_2=\alpha_6-\alpha_5$. In the vicinity of phase transition, 
the viscous moduli $\alpha_i$ have different dependences upon the order parameter 
$Q$: $\alpha_1\propto Q^2, \alpha_2,\alpha_3,\alpha_5, \alpha_6\propto Q, 
\alpha_4\propto Q^0$ \cite{Genn93}. 

Tables 1 and 2 (see Appendix) summarize viscosities and other 
physical parameters that characterize the most frequently used and well 
studied nematic LC, also known as {\it MBBA} and {\it PAA}.
\subsection{Basic equations}
\label{equat}
The complete system of hydrodynamic equations for nematic LC reflect 
the conservation laws of mass, and of linear and angular momenta.

\noindent
1. Continuity equation
\begin{eqnarray}
\frac{\partial \rho}{\partial t}+{\rm div}\left(\rho{\bf V}\right)=0\;.
\label{contin1} 
\end{eqnarray}
2. Navier--Stokes equation for visco--elastic LC
\begin{eqnarray}
\rho \frac{\partial V_i}{\partial t}+
\rho \langle {\bf V},\nabla_3\rangle V_i=\frac{\partial }{\partial x_k}
\left(T_{ik}^{\sf r}+T_{ik}^{\sf d}\right)\;.
\label{contin2}
\end{eqnarray}
3. Lesli--Ericksen equation of angular motion of the director ${\bf n}({\bf r},t)$
\begin{eqnarray}
\frac{\partial {\bf n}}{\partial t}={\bf m}^{\sf r}+{\bf m}^{\sf d}\;.   
\label{contin3}
\end{eqnarray}
The last equation is written for a negligible specific angular moment of inertia 
${\cal J}_{LC}$ of the LC, namely, ${\cal J}_{LC}\ll \rho\;r_0^2$, where $r_0$ is a 
characteristic size of the system. This is true in our case, where $r_0$ denotes  
radius of the jet.

Consider an isothermal incompressible jet, flowing along the $z$ axis, out of a nozzle at 
a velocity ${\bf V}$. The initial orientation of director ${\bf n}^0$ is assumed 
collinear with ${\bf V}$. The deviation from initial values of the director and 
pressure are defined as ${\bf n}^1={\bf n}-{\bf n}^0$, and $P_1=P-P_0$ 
respectively, where $P_0=\sigma/r_0$ is the unperturbed pressure within the 
cylindrical jet. Applying the linear approximation $|{\bf n}^1|\ll 1$,
equations (\ref{contin1})--(\ref{contin3}) are simplified as follows
\begin{eqnarray}
&&{\rm div}{\bf V}=0\;,\;\;\;\;
\rho\frac{\partial V_i}{\partial t}=-\frac{\partial P_1}{\partial x_i}+
\frac{\partial T_{ik}^{\sf din}}{\partial x_k}+
\frac{1-\lambda}{2}n^0_i\;{\rm div}{\bf F}-
\frac{1+\lambda}{2}\langle{\bf n}^0,\nabla_3\rangle F_i\;,\nonumber\\
&&\frac{\partial n^1_i}{\partial t}=n^0_k\Omega_{ki}+\lambda\;
\xi^0_{ij}\Upsilon_{jk}n^0_k+\frac{1}{\gamma_1}F_i\;,\;\;\;\;\;
\xi^0_{ij}=\delta_{ij}-n^0_in^0_j\;,\;\;\;i,j,k=x,y,z\;.
\label{contin4}
\end{eqnarray}
Choosing $n^0_z=1$, gives $F_z=0$ and hence 
\begin{eqnarray}
0&=&\frac{\partial V_x}{\partial x}+
\frac{\partial V_y}{\partial y}+\frac{\partial V_z}{\partial z}\;,\label{contin6}\\
\rho\frac{\partial V_x}{\partial t}&=&-\frac{\partial P_1}{\partial x}+
\left[\beta_1\;\Delta_2+\beta_2\frac{\partial^2}{\partial z^2}\right]V_x+
\left(\beta_2-\beta_1\right)
\frac{\partial^2 V_z}{\partial x \partial z}-
\frac{\lambda+1}{2}\frac{\partial F_x}{\partial z}\;,\label{contin6a}\\
\rho\frac{\partial V_y}{\partial t}&=&-\frac{\partial P_1}{\partial y}+
\left[\beta_1\;\Delta_2+\beta_2\frac{\partial^2}{\partial z^2}\right]V_y+
\left(\beta_2-\beta_1\right)\frac{\partial^2 V_z}{\partial y \partial z}-
\frac{\lambda+1}{2}\frac{\partial F_y}{\partial z}\;,\nonumber\\
\rho\frac{\partial V_z}{\partial t}&=&-\frac{\partial P_1}{\partial z}+
\left[\beta_2\;\Delta_2+\beta_3\frac{\partial^2}{\partial z^2}\right]V_z-
\frac{\lambda-1}{2}\;{\rm div}{\bf F}\;,\nonumber\\
\frac{\partial n^1_x}{\partial t}&=&
\frac{\lambda+1}{2}\frac{\partial V_x}{\partial z}+
\frac{\lambda-1}{2}\frac{\partial V_z}{\partial x}+\frac{F_x}{\gamma_1}\;,
\label{contin6b}\\
\frac{\partial n^1_y}{\partial t}&=&
\frac{\lambda+1}{2}\frac{\partial V_y}{\partial z}+
\frac{\lambda-1}{2}\frac{\partial V_z}{\partial y}+
\frac{F_y}{\gamma_1}\;,\;\;\;\;\;\;
\frac{\partial n^1_z}{\partial t}=0\;,\nonumber
\end{eqnarray}
where $\Delta_2=\frac{\partial^2}{\partial x^2}+\frac{\partial^2}{\partial y^2}$ is 
the two--dimensional Laplacian, $\beta_1=\eta_1^{\sf in},\;\beta_2=\eta_3^{\sf 
in}/2,\;\beta_3=\eta_5^{\sf in}-\eta_3^{\sf in}/2$ and $F_x,F_y$ are given in 
(\ref{frank3b}).  As isotropic viscosity means $\beta_i=\beta$, the above  
mentioned liquid crystals, {\it MBBA} and {\it PAA}, are clearly far from 
being isotropic (see Tables 1, 2 in Appendix).

In order to make the problem more specific and easier to solve, we consider 
{\it axisymmetrical} disturbances in a system of cylindrical LC jet, with radius 
$r_0$ and subject to the single elastic approximation ($K_i=K$). This provides the 
simplest approximation which still preserves the influence of elastic forces, 
on the hydrodynamics of an incompressible and elastic LC. In this case 
\begin{eqnarray}
0&=&\frac{\partial V_z}{\partial z}+\frac{\partial V_r}{\partial r}+
\frac{V_r}{r}\;,\label{contin7d}\\
\rho\frac{\partial V_r}{\partial t}&=&-\frac{\partial P_1}{\partial r}+
\left[\beta_1\left(\Delta_{2c}-\frac{1}{r^2}\right)+\beta_2\frac{\partial^2 }
{\partial z^2}\right]V_r+
\left(\beta_2-\beta_1\right)\frac{\partial^2 V_z}{\partial r \partial z}
-\mu_1 \frac{\partial F_r}{\partial z}\;,\label{contin7a}\\
\rho\frac{\partial V_z}{\partial t}&=&-\frac{\partial P_1}{\partial z}+
\left[\beta_2\Delta_{2c}+\beta_3\frac{\partial^2 }{\partial z^2}\right]V_z
-\mu_2\left(\frac{\partial F_r}{\partial r}+
\frac{F_r}{r}\right)\;,\label{contin7b}\\
\gamma_1\frac{\partial n^1_r}{\partial t}&=&\gamma_1 \mu_1\frac{\partial 
V_r}{\partial z}+\gamma_1\mu_2\frac{\partial V_z}{\partial r}+F_r\;,
\;\;\;\;n^1_z=0\;,\label{contin7c}
\end{eqnarray}
where
\begin{eqnarray}
\Delta_{2c} =\frac{\partial^2 }{\partial r^2}+
\frac{1}{r}\frac{\partial }{\partial r}\;,\;\;\;\;
F_r=K\left(\Delta_{2c}-\frac{1}{r^2}+\frac{\partial^2 }
{\partial z^2}\right)n^1_r\;,\;\;\;\;
\mu_1=\frac{\lambda+1}{2}\;,\;\;\mu_2=\frac{\lambda-1}{2}\;.
\label{contin7e}
\end{eqnarray}

Equations (\ref{contin7d})--(\ref{contin7c}) describe ordinary linear hydrodynamic 
behaviour of isotropic incompressible liquids provided that the LC properties 
vanish: $K,\gamma_1\rightarrow 0$ and $\beta_i=\beta$. The result is the well 
known continuity and linearized Navier--Stockes equations.
\begin{eqnarray}
{\rm div}{\bf V}=0\;,\;\;\;
\rho\frac{\partial {\bf V}}{\partial t}=-\nabla P_1+\beta \Delta_3 {\bf V}\;.
\label{contin8}
\end{eqnarray}

\subsection{Boundary conditions at free surface}
\label{boundary}
Boundary conditions at the free surface of a liquid crystal state that the jump in 
normal stress consists of two parts: one depends on the surface tension $\sigma$, and 
the other on the elastic disturbance $W_{elast}$ of the uniform director field 
${\bf n}_0({\bf r})$. Assuming that no tangential stresses exist at the free 
surface, the boundary conditions can be expressed as,
\begin{eqnarray}
\left(T_{ik}^{\sf r}+T_{ik}^{\sf din}\right)e_k+
\left(2 \sigma{\cal H}+W_{elast}\right) e_i+
\frac{\partial \sigma}{\partial x_i}=0\;\;\;\mbox{at}\;\;\;r=r_0\;,
\label{bound1}
\end{eqnarray}
where $e_i$ are the components of the normal unit vector {\bf e} in the 
reference frame of the LC--cylinder, and ${\cal H}=1/2\left(1/R_1+1/R_2\right)$ 
denotes mean surface curvature with principal radii $R_1$ and $R_2$.

The non--hydrodynamic part of boundary conditions at the free surface holds,  
provided that the scale of deformation of the initial surface is considerably 
larger compared to the molecular length of LCs 
\footnote{Strictly speaking, this assumption is correct when an equilibrium 
distribution of director field ${\bf n}({\bf r})$ is free of singularities. 
The problem of {\it minimal surface of LC drop} presents another situation 
wherein an essential rearrangement of the field ${\bf n}({\bf r})$, at the 
surface, can diminish the total energy by destroying the disclination core 
within the drop.}. 
This dictates a tangential behaviour of a smoothly disturbed director {\bf n} 
at the free surface, $e_z\ll e_r\sim 1$ :
\begin{equation}
\langle {\bf e}, {\bf n}\rangle=0\;\;\;\rightarrow\;\;e_z+n_r^1=0\;
\;\;\mbox{at}\;\;\;r=r_0\;.
\label{bound2}
\end{equation}
The last constraint cancels the gradient term in (\ref{bound1}). Finally we come  
to the boundary conditions in the linear approximation of the variables 
$n_r^1,V_r,V_z,P_1$ 
\begin{eqnarray}
T_{rr}^{\sf r}+T_{rr}^{\sf din}+2 \sigma{\cal H}+W_{elast}=0,\;\;\;
T_{zr}^{\sf r}+T_{zr}^{\sf d}=0\;.
\label{bound3}
\end{eqnarray}
Substitution of the expressions for the reactive and dissipative stress tensors  
gives
\begin{eqnarray}
2\beta_1 \Upsilon_{rr}-P_1=2 \sigma_0\left({\cal H}_0-{\cal H}\right)
-W_{elast},\;\;\;
2\beta_2 \Upsilon_{zr}=\mu_2F_r\;\;\;\mbox{at}\;\;\;r=r_0\;.
\label{bound4}
\end{eqnarray}
where ${\cal H}_0=1/2r_0$ is the initial mean curvature of the LC--cylinder. 
The equations for a jet surface, disturbed by a wave $\zeta(z,t)$, 
and its radial velocity $\partial \zeta/\partial t$, are given by
\begin{equation}
r(z,t)=r_0+\zeta(z,t)\;,\;\;\;V_r=\frac{\partial \zeta}{\partial t}
\;\;\;\mbox{at}\;\;\;r=r_0\;,
\label{bound5}
\end{equation}
where $\zeta\ll r_0$ is the radial displacement of a surface point. 
The principal radii of the surface curvature, in the context of linear 
approximation with respect to $\zeta$, and its derivatives can be expressed as, 
\begin{equation}
\frac{1}{R_1}=\frac{1}{r_0+\zeta}\cong 
\frac{1}{r_0}-\frac{\zeta}{r_0^2}\;,\;\;\;
\frac{1}{R_2}\cong -\frac{\partial^2 \zeta}{\partial z^2}\;.
\label{bound5a}
\end{equation}
This transforms the boundary conditions (\ref{bound2}), (\ref{bound4}) into
\begin{eqnarray}
n_r^1&=&\frac{\partial \zeta}{\partial z}\;,\;\;\;\;
V_r=\frac{\partial \zeta}{\partial t}\;,
\label{bound6a}\\
2\beta_2 \Upsilon_{zr}&=&\mu_2F_r\;,\label{bound6b}\\
P_1-2\beta_1 \Upsilon_{rr}&=&-\sigma_0\left(
\frac{\zeta}{r_0^2}+\frac{\partial^2 \zeta}{\partial z^2}\right)+
W_{elast}\;.
\label{bound6c}
\end{eqnarray}
The term $W_{elast}$ deserves further discussion. It reflects 
the existence of normal stresses, at the surface, which arise due to the 
resistance of the uniformly orientated continuos LC media to a surface 
disturbance. $W_{elast}$ vanishes in undisturbed LC jets and it depends 
linearly on the elastic modulus $K$, radius $r_0$ and derivatives 
of $\zeta$. Moreover, an invariance of the problem with respect to inversion 
of the $z$--axis requires sole dependence on derivatives of even orders. 
An explicit expression for $W_{elast}$ is derived in Section \ref{contr}.

\section{Plateau instability in a LC cylinder}
\label{plateau1}
Before proceeding to tackle the sophisticated mathematics of equations 
(\ref{contin7d})--(\ref{contin7c}), as supplemented by boundary conditions
(\ref{bound6a})--(\ref{bound6c}), capillary instability of the LC cylinder is
discussed. This is done by applying the Plateau considerations \cite{Plat73} on the 
figures of a liquid mass withdrawn from the action of gravity.

Consider a LC cylinder with a surface disturbed as specified by (\ref{bound5}), 
where $\zeta=\zeta_0 \cos kz$, $\zeta_0$ is small compared to $r_0$, and 
$k=2\pi/\Lambda$, $\Lambda$ being the disturbance {\it wavelength}. The idea of 
Plateau, applied here, is to find such {\it cut--off wavelength} $\Lambda_s$ 
of the disturbance, that defines breakage of the cylinder into droplets 
with due decrease of the total energy.

The volume $v$ enclosed within one wave length is given by
\begin{equation}
v=\int_v dv=\pi \left(r_0^2+\frac{1}{2} \zeta_0^2\right)\;\;\;\rightarrow
\;\;\;r_0=\sqrt{\frac{v}{\pi}}\left(1-\frac{1}{4}\frac{\pi\zeta_0^2}{v}
\right)\;,
\label{plat1}
\end{equation}
where $r_0$ in the right h.s. of (\ref{plat1}) is given as a second order expansion 
$\zeta_0$. 
The total energy ${\cal E}$ of the LC cylinder with a disturbed director field 
${\bf n}({\bf r})$ is given by
\begin{eqnarray}
{\cal E}=\sigma_0 \int_s ds+\frac{K}{2}\int_v 
\left({\rm div}^2{\bf n}+{\rm rot}^2{\bf n}\right)\;dv\;.\label{plat2}
\end{eqnarray}
The static director field ${\bf n}({\bf r})$ can be found from equation 
(\ref{contin7e}) and the attendant boundary condition (\ref{bound6a})
\begin{eqnarray}
n_z^0=1\;,\;\;F_r=0\;\;\;\rightarrow\;\;\;
\left(\Delta_{2c}-\frac{1}{r^2}+\frac{\partial^2 }
{\partial z^2}\right)n^1_r=0\;,\;\;
n_r^1=\frac{\partial \zeta}{\partial z}\;\;\;\mbox{at}\;\;\;r=r_0\;.
\label{plat2a}
\end{eqnarray}
Equation (\ref{plat2a}) is satisfied by the following solution, 
which is finite at $r=0$
\begin{equation}
n_r^1(r,z)=-\frac{k\zeta_0}{I_1(kr_0)}I_1(kr)\;\sin kz\;,
\label{plat2b}
\end{equation}
where $I_m(x)$ is a modified Bessel function of order $m$. 
The contribution of elastic forces is determined by
\begin{equation}
{\rm div}^2{\bf n}+{\rm rot}^2{\bf n}=
k^2\left[\frac{k\zeta_0}{I_1(kr_0)}\right]^2
\left[A_1^2(kr)\sin^2 kz+A_2^2(kr)\cos^2 kz\right]
\label{plat2c}
\end{equation}
where
$$
A_1(y)=\frac{dI_1(y)}{dy}+\frac{1}{y}I_1(y)\;,\;\;
A_2(y)=I_1(y)\;.
$$
A simple integration of (\ref{plat2}) gives
\begin{equation}
{\cal E}=2 \pi \sigma_0 r_0 \left(1+\frac{1}{4} k^2\zeta_0^2\right)+
\frac{\pi}{2} K \left[\frac{k \zeta_0}{I_1(kr_0)}\right]^2
\int_0^{kr_0}\left[A_1^2(y)+A_2^2(y)\right]ydy\;.
\label{plat2d}
\end{equation}
Inserting $r_0$ from (\ref{plat1}) into the first term above, we obtain
\begin{equation}
{\cal E}-2 \sigma_0\sqrt{\pi v}=\sigma_0\frac{\pi \zeta_0^2}{2 r_0}
\left(\varpi^2-1\right)+
\frac{\pi}{2} K \left[\frac{\zeta_0\varpi}{r_0 I_1(\varpi)}\right]^2
\int_0^{\varpi}\left[A_1^2(y)+A_2^2(y)\right]ydy\;,\;\;\;\varpi=kr_0.
\label{plat2e}
\end{equation}
The positive root $\varpi_s=k_sr_0$, given in the right h.s. of (\ref{plat2e}), 
determines the cut--off  wavelength $\Lambda_s$ of capillary disturbances,
which renders the LC cylinder unstable. Subsequent disintegration into 
detached masses is favored by the decrease in ${\cal E}$ 
\begin{equation}
\left(\varpi_s^2-1\right)+\varkappa \frac{\varpi_s^2}{I_1^2(\varpi_s)}
\int_0^{\varpi_s}\left[A_1^2(y)+A_2^2(y)\right]ydy=0\;,
\;\;\;\;\varkappa=\frac{K}{\sigma_0r_0}\;,
\label{plat2f}
\end{equation}
where the subscript $"s"$ denotes the static nature of Plateau 
instability. 

\begin{figure}[h]
\centerline{\psfig{figure=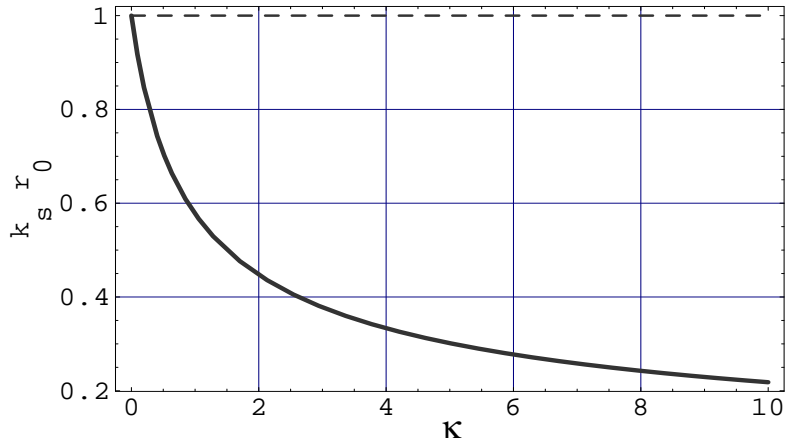,height=6cm,width=11cm}}
\vspace{-.1cm}
\caption{Universal plots of $k_sr_0$ {\it vs.} $\varkappa$ for Plateau 
instabilities in LC cylinder ({\it plain line}), and in ordinary liquid 
$k_sr_0=1$ ({\it dashed line}). }
\label{figplat1}
\end{figure}

The quadratic approximation (\ref{frank1}) with respect to the derivatives
$\partial {\bf n}/\partial x_j$, which provides the basis for the Frank theory,  makes 
the expression (\ref{plat2f}) correct only in terms of the $\varpi_s^2$ approximation. 
Indeed, the power of $\varpi_s$ in (\ref{plat2f}) should not exceed 2, otherwise the 
calculation becomes inconsistent. Thus, we get
\begin{equation}
{\cal E}-2 \sigma_0\sqrt{\pi v}=\sigma_0\frac{\pi \zeta_0^2}{2 r_0}
\left(\varpi^2-1\right)+\pi K k^2 \zeta_0^2\;\;\;\;\mbox{and}\;\;\;\;
\varpi_s=\frac{1}{\sqrt{1+2\varkappa}}\;.
\label{platt1}
\end{equation}
The asymptotic behaviour of $\varpi_s(\varkappa)$ shows two important limits: 
\begin{eqnarray}
\varpi_s=1-\varkappa\;\;\;\mbox{if}\;\;\;\varkappa\ll 1\;;\;\;\;
\varpi_s=\frac{1}{\sqrt{2\varkappa}}\left(1-\frac{1}{4\varkappa}\right)
\;\;\;\mbox{if}\;\;\;\varkappa\gg 1\;.
\label{plat2g}
\end{eqnarray}
Figure \ref{figplat1} shows a plot of $k_sr_0$ {\it vs.} $\varkappa$ 
for Plateau instabilities in LC and in ordinary liquid.

The corresponding asymptotic cut--off wavelength $\Lambda_s$ are obtained as
\begin{eqnarray}
\Lambda_s=2\pi r_0\left(1+\varkappa\right)\;\;\;
\mbox{if}\;\;\;\varkappa\ll 1\;;\;\;\;
\Lambda_s=2\pi\sqrt{\frac{2 K}{\sigma_0}}\;\sqrt{r_0}
\left(1+\frac{1}{4\varkappa}\right)
\;\;\;\mbox{if}\;\;\;\varkappa\gg 1\;.
\label{plat2gg}
\end{eqnarray}
This result shows that $k\geq k_s$ increases the total energy ${\cal E}$ of the 
disturbed system, whereas $k\leq k_s$ decreases it. According to (\ref{plat2g}), there 
are two marginal regimes of instability:

$\bullet$ {\it Capillary regime} $r_0\gg K/\sigma_0$. Here $\Lambda_s$
is close in value to the circumference of the cylinder and the elastic 
deformation contribution $\int E_d dv$, to the total energy ${\cal E}$, is 
negligible.

\noindent
This regime must apply to a wide range of nematic LC, 
since the common values of $K\simeq 10^{-11}J/m$ \cite{Genn93} and
$\sigma_0\simeq 10^{-2}J/m^2$ \cite{Cogn82} lead to 
$K/\sigma_0\simeq 10^{-9}m$. This value is evidently smaller than the
presently attainable radii of the jet.

$\bullet$ {\it Elastic  regime} $r_0\ll K/\sigma_0$. This case reflects the 
dominance of elastic deformation and predicts an unusual behaviour for 
$\Lambda_s\sim \sqrt{r_0}$.

\noindent
This regime cannot be reached by simple increase of the elastic moduli since 
their magnitude is determined by $K\sim \kappa T/a$, where $\kappa T\sim 
10^{-20} J$ is the Bolzmann thermal energy at room temperature, and $a\sim 
10^{-9}m$ denotes molecular length of LC. In contrast, the effect of surface 
tension can be diminished by surfractants or by 
charging the surface of the liquid. In the latter case the charge can 
virtually eliminate the effect of surface tension and provide the 
conditions where the elastic forces predominate.

\subsection{$W_{elast}$ and Gaussian surface curvature}
\label{contr}
The straightforward way to derive an expression for $W_{elast}$ is 
to solve the elastic problem for the stresses existing on a deformed axisymmetric  
surface of a LC cylinder. This relates to the Plateau instability, which 
obviates the need to repeat the entire procedure.

When we turn from Plateau considerations on the {\it static} instability
of LC cylinders to the capillary instability of LC jets, the question is 
whether the cut--off wavelengths of both the static $\Lambda_s$ and 
hydrodynamic $\Lambda_d$ problems coincide. This question was skipped by 
Rayleigh in his studies of isotropic viscous liquids, since for ordinary 
liquids both cut--off wavelengths always coincide $\Lambda_s\equiv\Lambda_d$. 
This identity reflects a deep equivalence principle of the bifurcation point for 
non--trivial steady state of dynamic system, and the threshold of static 
instability concerned with a minimum of its free energy ${\cal E}$ \cite{Chandra61}.

Making use of $\Lambda_s\equiv\Lambda_d$ we construct the term 
$W_{elast}$ which enters the boundary condition (\ref{bound6c}). To this 
end, we examine and represent the total energy (\ref{platt1}) as follows
\begin{eqnarray}
{\cal E}-2 \sigma_0\sqrt{\pi v}=\frac{\pi \zeta_0 r_0}{2}
\left[-\sigma_0\left(\frac{\zeta_0}{r_0^2}-\zeta_0 k^2\right)+
2K\frac{\zeta_0}{r_0}k^2\right]\;.
\label{contrib1}
\end{eqnarray}
Next, we compare the expression within the brackets with the right h.s. of 
(\ref{bound6c}). This gives $W_{elast}$, which generates 
the elastic contribution in (\ref{contrib1})
\begin{equation}
W_{elast}=2K{\cal G}\;,\;\;\;\;
{\cal G}=\frac{1}{R_1R_2}=
-\frac{1}{r_0}\frac{\partial^2 \zeta}{\partial z^2}\;,
\label{contrib2}
\end{equation}
where ${\cal G}$ is the Gaussian surface curvature in accordance with 
(\ref{bound5a}). Thus the final expression for 
boundary conditions (\ref{bound1}) is based on two fundamental 
invariants of the surface curvature, i.e. mean surface curvature ${\cal H}$, 
and Gaussian surface curvature ${\cal G}$.

\section{Dispersion relation}
\label{dispers}
Rayleigh was the first to observe \cite{Rayleigh79} that contrary to Plateau, 
the instability problem is not so definite. The mode whereby a system 
deviates from 
unstable equilibrium must depend on the nature and characteristics of the small 
displacements to which this system is subjected. In the absence of such displacement, 
any system, however unstable, cannot depart from equilibrium. 
These characteristics, being hydrodynamic, reflect the effect of viscosity, 
which predominates over that of inertia. In the case of ordinary liquids, 
the mode of {\it maximum} instability which corresponds to the wavelength 
$\Lambda_R=4.508\times 2r_0$ exceeds the circumference of the liquid 
cylinder. We anticipate that the instability of LC jets possesses similar 
features.

The fact that a velocity potential does not exist in an anisotropic visco--elastic 
liquid, dictates a standard approach to this problem which was elaborated first 
by Rayleigh \cite{Rayleigh92}. Let us define the {\it Stokes stream function} 
$\Psi({\bf r},t)$ and a {\it director\;potential} $\Theta({\bf r},t)$ as, 
\begin{equation}
V_r=-\frac{1}{r}\frac{\partial \Psi}{\partial z}\;,\;\;
V_z=\frac{1}{r}\frac{\partial \Psi}{\partial r}\;\;\;\;\mbox{and}\;\;\;
n^1_r=\frac{\partial \Theta}{\partial r}\;,
\label{disper2}
\end{equation}
so that the continuity equation (\ref{contin7d}) holds. From the other three
equations (\ref{contin7a})--(\ref{contin7c}) we have
\begin{eqnarray}
\frac{\partial P_1}{\partial r}&=&
\left(\beta_2-\beta_1\right)\frac{\partial^2 }{\partial r \partial z}
\left(\frac{1}{r}\frac{\partial \Psi}{\partial r}\right)
-\frac{1}{r}\frac{\partial }{\partial z}
\left[\beta_1r\frac{\partial }{\partial r}
\left(\frac{1}{r}\frac{\partial \Psi}{\partial r}\right)+
\beta_2\frac{\partial^2 \Psi}{\partial z^2}-\rho\frac{\partial\Psi}{\partial t}
+\mu_1 r F_r\right]\;\;\;
\label{disper3b}\\
\frac{\partial P_1}{\partial z}&=&\frac{1}{r}\frac{\partial }{\partial r}
\left[\beta_2 r\frac{\partial }{\partial r}
\left(\frac{1}{r}\frac{\partial \Psi}{\partial r}\right)+
\beta_3\frac{\partial^2 \Psi}{\partial z^2}-
\rho\frac{\partial\Psi}{\partial t}-\mu_2 r F_r\right]\;,
\label{disper3c}\\
\frac{\partial^2 \Theta}{\partial r \partial t}&=&\frac{1}{r}
\left[\mu_2 r \frac{\partial }{\partial r}\left(\frac{1}{r}
\frac{\partial \Psi}{\partial r}\right)-\mu_1
\frac{\partial^2 \Psi}{\partial z^2}\right]+\frac{1}{\gamma_1}F_r\;,\;\;\;\;
F_r=K \left(\Delta_{2c}+\frac{\partial^2 }{\partial z^2}-
\frac{1}{r^2}\right) \frac{\partial 
\Theta}{\partial r}\;.\;\;
\label{disper3d}
\end{eqnarray}
Applying the commutation rules give, 
$$
\left(\Delta_{2c}-\frac{1}{r^2}\right)\frac{\partial \Theta}{\partial r}=
\frac{\partial }{\partial r}\Delta_{2c} \Theta\;\;\;
\rightarrow\;\;\;
F_r=K \frac{\partial }{\partial r}\left(\Delta_{2c}+\frac{\partial^2 
}{\partial z^2}\right)\Theta\;,
$$
which facilities simplification of the above equations. Assuming that an 
axisymmetrical disturbance, characterized by a wavelength $2\pi/k$, increases 
exponentially in time with the growth rate $s$, gives,
\begin{equation}
\left\{\Psi,\;\Theta,\;\zeta,\;P_1,\;F_r\right\}=
\left\{i\;\psi(r),\;i\;\theta(r),\;\varsigma(r),
\;p(r),\;i\;f(r)\right\}\times e^{st+ikz}\;,
\label{disper4}
\end{equation}
Inserting (\ref{disper4}) into (\ref{disper3b})--(\ref{disper3d}) 
gives rise to the following amplitude equations
\begin{eqnarray}
\frac{1}{k}\frac{\partial p}{\partial r}&=&
\beta_4\frac{\partial }{\partial r}
\left(\frac{1}{r}\frac{\partial \psi}{\partial r}\right)-
(\beta_2k^2+s\rho)\frac{\psi}{r}+\mu_1 f\;,\;\;\;\;\;\;
\beta_4=2\beta_1-\beta_2\;,\label{disper5a}\\
kp&=&\frac{1}{r}\frac{\partial }{\partial r}\left\{r
\left[\beta_2\frac{\partial }{\partial r}
\left(\frac{1}{r}\frac{\partial \psi}{\partial r}\right)-
(\beta_3k^2+s\rho)\frac{\psi}{r}-\mu_2 f\right]\right\}
\;,\;\label{disper5b}\\
s\frac{\partial \theta}{\partial r}&=&
\mu_2 \frac{\partial }{\partial r}\left(\frac{1}{r}
\frac{\partial \psi}{\partial r}\right)+\mu_1 k^2 \frac{\psi}{r}+
\frac{1}{\gamma_1}f\;,\;\;\;\;\;
f=K\frac{\partial }{\partial r}\left(\Delta_{2c}-k^2\right)\theta\;,
\label{disper5d}
\end{eqnarray}

The new variables in (\ref{disper4}) require reformulation of the boundary
conditions (\ref{bound6a})--(\ref{bound6c}) as follows, 
\begin{eqnarray}
k\varsigma=\frac{\partial \theta}{\partial r}\;,\;\;\;\;\;
s\varsigma=k\frac{\psi}{r}\;,\;\;\;\;\;
\frac{\mu_2}{\beta_2}\;f=
\frac{\partial }{\partial r}\left(\frac{1}{r}
\frac{\partial \psi}{\partial r}\right)+k^2\frac{\psi}{r}\;,\;\;\;\;\;
p&=&2\beta_1k\frac{\partial }{\partial r}\left(\frac{\psi}{r}\right)+
\varsigma \Gamma\;
\label{disper5e}
\end{eqnarray}
where
$$
\Gamma=\sigma_0 \left(k^2-\frac{1}{r_0^2}\right)+
2K\frac{1}{r_0}k^2\;.
$$
The real form of the amplitude equations (\ref{disper5a})--(\ref{disper5d})
and boundary conditions (\ref{disper5e}) imply that (\ref{disper4}) divides  
the five variables into two groups: $P_1,\zeta$ and $\Psi,\Theta,F_r$. These 
groups are shifted with respect to each other by the phase angle $\pi/2$. 
\subsection{Reduction of the amplitude equations}
\label{cut1}
In this Section we perform a standard procedure for the simplification of the 
amplitude equations (\ref{disper5a})--(\ref{disper5d}). Substituting $f$ 
from (\ref{disper5d}), into the other amplitude equations we get
\begin{eqnarray}
\frac{1}{k}\frac{\partial p}{\partial r}&=&
B_1\frac{\partial }{\partial r}\left(\frac{1}{r}\frac{\partial \psi}
{\partial r}\right)-(B_2k^2+s\rho)\frac{\psi}{r}+
s\gamma_1\mu_1\frac{\partial \theta}{\partial r}\;,\label{disper5f}\\
kp&=&\frac{1}{r}\frac{\partial }{\partial 
r}\left\{r\left[B_3\frac{\partial }
{\partial r}\left(\frac{1}{r}\frac{\partial \psi}{\partial r}\right)
-(B_4k^2+s\rho)\frac{\psi}{r}\right]\right\}-
s\gamma_1\mu_2 \frac{1}{r}\frac{\partial }{\partial r}
\left(r\frac{\partial \theta}{\partial r}\right)\;,
\label{disper5x}\\
0&=&\mu_2 \frac{\partial }{\partial r}\left(\frac{1}{r}
\frac{\partial \psi}{\partial r}\right)+\mu_1 k^2 \frac{\psi}{r}+
\frac{K}{\gamma_1}\frac{\partial }{\partial r}\left[
\frac{1}{r}\frac{\partial }{\partial r}
\left(r\frac{\partial \theta}{\partial r}\right)-
\left(k^2+\frac{s\gamma_1}{K}\right)\theta\right]\;,
\label{disper5q}
\end{eqnarray}
where
\begin{eqnarray}
B_1=\beta_4-\gamma_1\mu_1\mu_2\;,\;\;B_2=\beta_2+\gamma_1\mu_1^2\;,\;\;
B_3=\beta_2+\gamma_1\mu_2^2\;,\;\;
B_4=\beta_3-\gamma_1\mu_1\mu_2\;.\label{disp1a}
\end{eqnarray}
and $B_2>0,B_3>0$ by virtue of (\ref{lesli1}). Let a new stream function 
$\chi$ be defined as $\psi=r\;\partial \chi/\partial r$. The orientational
$\vartheta$ and kinematic $\nu_i$ viscosities, as well as other auxiliary 
functions, are defined by the following relations
\begin{eqnarray}
\vartheta=\frac{K}{\gamma_1}\;,\;\;\nu_i=\frac{B_i}{\rho}\;,\;\;
u_i^2=k^2+\frac{s}{\nu_i}\;,\;\;w^2=k^2+\frac{s}{\vartheta}\;,
\;\;\;\;\;\frac{\vartheta}{\nu_i}\ll 1\;\rightarrow\;
u_i^2\leq w^2\;,
\label{disp1b}
\end{eqnarray}
where the first inequality in (\ref{disp1b}) applies to known nematic LC fluids 
(see Tables 1, 2 in Appendix). Using the new notations we 
find the first integrals of the amplitude equations as, 
\begin{eqnarray}
\frac{p}{k}&=&\left(B_1\Delta_{2c}-B_2u_2^2\right)\chi+
s\gamma_1\mu_1\theta\;,\label{disper6a}\\
kp&=&\left(B_3\Delta_{2c}-B_4u_4^2\right)\Delta_{2c}\chi-
s\gamma_1\mu_2\Delta_{2c}\theta\;,\label{disper6b}\\
0&=&\left(\mu_2\Delta_{2c}+\mu_1k^2\right)\chi+
\vartheta\left(\Delta_{2c}-w^2\right)\theta\;.\label{disper6c}
\end{eqnarray}
Next, we eliminate the pressure amplitude $p$ from (\ref{disper6a}) and 
(\ref{disper6b}). This gives,
\begin{eqnarray}
\left[B_3\Delta_{2c}^2-\left(B_1k^2+B_4u_4^2\right)\Delta_{2c}+
B_2u_2^2 k^2\right]\chi-s\gamma_1\left(\mu_2\Delta_{2c}+
\mu_1 k^2\right)\theta&=&0\;,\label{disper7a}\\
\left(\mu_2\Delta_{2c}+\mu_1 k^2\right)\chi+
\vartheta\left(\Delta_{2c}-w^2\right)\theta&=&0\;.\label{disper7b}
\end{eqnarray}
Diagonalizing a matrix of operators in (\ref{disper7a}) and (\ref{disper7b}) we 
obtain the following homogeneous equations for the functions $\chi(r)$ and 
$\theta(r)$, 
\begin{eqnarray}
\left[D_3\Delta_{2c}^3-D_2\Delta_{2c}^2+D_1\Delta_{2c}-D_0\right]
\left(\begin{array}{c}
\chi\\
\theta\end{array}\right)=
\left(\begin{array}{c}
0\\
0\end{array}\right)\;,
\label{disper8}
\end{eqnarray}
where
\begin{eqnarray}
D_0&=&k^2\left(\vartheta B_2 u_2^2 w^2-s\gamma_1\mu_1^2k^2\right)\;,\;\;
D_1=\vartheta \left(B_1k^2w^2+B_2 k^2u_2^2+B_4 w^2u_4^2\right)+
2s\gamma_1\mu_1\mu_2k^2\;,\;\nonumber\\
D_2&=&\vartheta\left(B_1k^2+B_3 w^2+B_4u_4^2\right)-s\gamma_1\mu_2^2\;,\;\;
D_3=\vartheta B_3\;.\label{disper8c}
\end{eqnarray}
It is easy to verify that all coefficients $D_j$ are positive, if the conditions 
that all $B_i>0$ and $\mu_2\ll 1$, $\vartheta/\nu_i\ll 1$ are satisfied. The 
latter are in a good agreement with numerous observations in nematic LCs 
\cite{Genn93}.

Further factorization (recalling that $D_3>0$) of the polynomial differential 
operator gives 
\begin{eqnarray}
D_3\Delta_{2c}^3-D_2\Delta_{2c}^2+D_1\Delta_{2c}-D_0=
D_3 \left(\Delta_{2c}-m_1^2\right)\left(\Delta_{2c}-m_2^2\right)
\left(\Delta_{2c}-m_3^2\right)\;.
\label{disp1d}
\end{eqnarray}
Equation (\ref{disp1d}) facilitates the following finite solutions of equations 
(\ref{disper8}) 
\begin{eqnarray}
\chi(r)=\sum_{j=1}^3\frac{C_j}{m_j}I_0(m_jr)\;,\;\;\;\;
\theta(r)=\sum_{j=1}^3\frac{G_j}{m_j}I_0(m_jr)\;,
\label{disper11}
\end{eqnarray}
where the second fundamental solutions that diverge at $r=0$ were excluded, 
$C_j$ and  $G_j$ are indeterminate coefficients and $m_j^2$ are three generic
\footnote{The freedom to choose the physical parameters of LC seems to
admit a degeneration of cubic equation (\ref{disper12}), when some of the 
roots $m_j^2$ can coincide in different ways. By virtue, such 
coincidence is not important, since it could occur only at specific 
wave vectors $k^*$, which the coefficients $D_2,D_1,D_0$ are dependent 
upon. By the other hand, this kind of degeneration might be interesting
if $k^*$ is accidentally close to the cut--off wave vector $k_d$, when the 
breakage of the LC jet develops.}
roots of the following cubic equation
\begin{eqnarray}
D_3m^6-D_2m^4+D_1m^2-D_0=0\;\;\rightarrow\;\;
\sum_{j=1}^3 m_j^2=\frac{D_2}{D_3},\;\;
\sum_{j\neq k}^3 m_j^2m_k^2=\frac{D_1}{D_3},\;\;
\prod_{j=1}^3 m_j^2=\frac{D_0}{D_3}\;.\;\;\;
\label{disper12}
\end{eqnarray}
The coefficients $G_j$ can be expressed through $C_j$, 
once (\ref{disper11}) is inserted into (\ref{disper7b}),
\begin{eqnarray}
G_j=\frac{1}{\vartheta}\;g_jC_j\;,\;\;\;g_j= 
\frac{\mu_1 k^2+\mu_2 m_j^2}{w^2-m_j^2}\;,\;\;\;\;j=1,2,3\;.
\label{disper13}
\end{eqnarray}
The amplitude of the pressure $p(r)$, the stream function $\psi(r)$ and 
the displacement of a point on the surface $\varsigma(r_0)$ are easily found 
from (\ref{disper5d}), (\ref{disper6a}), (\ref{disper7b}) and (\ref{disper13}), 
\begin{eqnarray}
p(r)&=&k\sum_{j=1}^3\frac{l_j}{m_j}C_jI_0(m_jr)\;,\;\;\;
l_j=B_1m_j^2-B_2u_2^2+\frac{s}{\vartheta}\;\gamma_1\mu_1g_j\;,\label{disper14a}\\
\psi(r)&=&r\sum_{j=1}^3C_jI_1(m_jr)\;,\;\;\;
\varsigma(r_0)=\frac{1}{\vartheta\;k}\sum_{j=1}^3g_jC_jI_1(m_jr_0)\;,
\;\;j=1,2,3\;.\nonumber
\end{eqnarray}
Before proceeding on to the end of this Section, we discuss the distribution of 
the roots $m_j^2$ of the cubic equation (\ref{disper12}) in the complex plane.

First, $m_1^2$ is always positive since $D_j>0$, as mentioned 
above, and following the Descartes' rule of signs interchange 
in the sequence of coefficients for real algebraic equations. The 
other two roots $m_{2,3}^2$ are either positive or complex--conjugate with
positive real parts. The last case leads in (\ref{disper11}) to 
Bessel functions of complex arguments. This fact can indicate that the 
separation of the two groups of functions $P_1,\zeta$ and $\Psi,\Theta,F_r$ by 
the $\pi/2$ phase angle, is more elaborate than assumed in 
(\ref{disper4}). Another consequence of the existence of complex--conjugated 
roots $m_j^2$, which is more important from the physical standpoint, is 
appearance of the imaginary contributions in the dispersion equation. This can 
lead to the complex value of the growth rate $s=\overline{s}+i\omega$, as its 
solution, and to the non--steady (oscillatory) evolution of the jet, e.g. 
$\zeta(z,t) \propto \varsigma(r_0) e^{\overline{s}t}\times e^{i(\omega t+kz)}$, 
where $\omega$ denotes frequency of oscillations. 

\subsection{Dispersion equation}
\label{revis}
In what follows we derive the dispersion equation $s=s(kr_0)$, which 
determines the evolution of Rayleigh instability in LC jets. The revised 
version of the boundary conditions (\ref{disper5e}) at $r=r_0$, which utilizes  
the new stream function $\chi(r)$, reads
\begin{eqnarray}
s\frac{\partial \theta}{\partial r}
=k^2\frac{\partial \chi}{\partial r}\;,\;\;\;\;
s\gamma_1\mu_2\frac{\partial \theta}{\partial r}
=B_3\frac{\partial }{\partial r}\Delta_{2c}\chi
+B_5k^2\frac{\partial \chi}{\partial r}\;,\;\;\;\;
\frac{s}{k}\;p=2s\beta_1 \frac{\partial^2 \chi}{\partial r^2}+ 
\Gamma\frac{\partial \chi}{\partial r}\;,
\label{revis1}
\end{eqnarray}
where $B_5=\beta_2+\gamma_1\mu_1\mu_2$. Substituting (\ref{disper11}) and 
(\ref{disper14a}) into (\ref{revis1}), and elimination of the coefficients 
$C_1,C_2,C_3$ from the linear equations, leads to a $(3\times 3)$--determinant 
equation
\begin{equation}
\det S_{ij}=0\;,
\label{revis2}
\end{equation}

\noindent
where 
\begin{eqnarray}
S_{1j}&=&k^2-\frac{s}{\vartheta}\;g_j,\;\;\;\;\;
S_{2j}=B_3m_j^2+B_5k^2-\frac{s}{\vartheta}\;\gamma_1\mu_2g_j\;,\nonumber\\
S_{3j}&=&\Gamma -s\left[\frac{l_j}{m_j}\frac{I_0(m_jr_0)}{I_1(m_jr_0)}-
2\beta_1m_j\frac{I_1'(m_jr_0)}{I_1(m_jr_0)}\right]\;,
\label{revis3}
\end{eqnarray}
and $I_1'(y)=dI_1(y)/dy$. Equation (\ref{revis2}) is an implicit form 
of the exact dispersion relation, which is highly complex and cannot be solved 
analytically in the general case. Nevertheless, here we can verify the fact, 
that the cut--off wavelength $\Lambda_d$ does coincide with $\Lambda_s$ obtained 
due to Plateau. Indeed, the cut--off regime corresponds to (\ref{revis1}) when 
$s=0$ and is satisfied for $\Gamma=0$, i.e. $\Lambda_d=\Lambda_s$. The implications 
of equation (\ref{revis2}) can be extended further: for the study of different 
modes of LC flow, including oscillations, and in order to describe asymptotic 
behaviour of LC jets. This is outside the scope of this paper and will be considered 
elsewhere. In the next Section, we present a case which facilitates decoupling
of hydrodynamic and orientational modes, and consequently the solution of the
Rayleigh instability problem in closed form.

\section{Decoupling of hydrodynamic and orientational modes}
\label{decoupl}
In this Section we discuss a case that renders the dispersion equation (\ref{revis2}) 
solvable. Here we encounter another problem: the elasticity of LC and anisotropy 
of its viscous properties have the same origin and therefore cannot be managed 
separately. Nevertheless, we consider the case where the dispersion equation 
(\ref{revis2}) can be simplified. The large number of physical parameters involved 
(three viscous moduli, two kinetic coefficients $\lambda$ and $\gamma_1$, 
orientational $\vartheta$ and kinematic viscosities $\nu_i$, and  
dimensionless parameter $\varkappa$) call for such a treatment.

This applies to LC with rod--like molecules ($\lambda\simeq 1$) and 
low orientational viscosity $\vartheta$ 
\begin{eqnarray}
\mu_1\simeq 1\;,\;\;\;\mu_2\simeq 0\;,\;\;\;\vartheta\ll \nu_i\;,\;\;\;
k^2\ll\frac{s}{\vartheta}\;,
\label{dec00}  
\end{eqnarray}
where the first three relations in (\ref{dec00}) apply to known nematic LC 
fluids (see Tables 1, 2 in Appendix). The last inequality in 
(\ref{dec00}) applies to the low--viscosity limit which was considered for 
the kinematic viscosity in ordinary liquids by Rayleigh \cite{Rayleigh79}.

In this case the characteristic equation (\ref{disper12}) reduces as follows
\begin{eqnarray}
m^6-\frac{s}{\vartheta}m^4+\frac{s}{\vartheta}\left({\cal B}k^2+
\frac{s}{\overline{\nu_2}}\right)m^2-\frac{s}{\vartheta}k^2\left(k^2+
\frac{s}{\overline{\nu_2}}\right)=0\;,\;\;\;
\overline{\nu_i}=\frac{\beta_i}{\rho}\;,\;\;\;
{\cal B}=\frac{\beta_3+\beta_4}{\beta_2}\;.
\label{dec01}
\end{eqnarray}
The three roots $m_j^2$ of equation (\ref{disper12}) read
\begin{eqnarray}
2m_{1,2}^2={\cal B}k^2+\frac{s}{\overline{\nu_2}}\pm
\sqrt{\left({\cal B}^2-4\right)k^4+
2 \left({\cal B}-2\right)k^2\frac{s}{\overline{\nu_2}}+
\left(\frac{s}{\overline{\nu_2}}\right)^2}\;,\;\;\;m_3^2=
\frac{s}{\vartheta}\;.
\label{dec001}
\end{eqnarray}
A simple analysis of (\ref{dec001}) shows that the dimensionless parameter ${\cal B}$ 
has a critical value 2 that separates two different evolution scenaria of the 
LC jet. If ${\cal B}>2$ then the both roots $m_1^2$ and $m_2^2$ 
are positive and the capillary instability always appears via trivial 
bifurcation (steady--state instability). This scenario applies to  
{\it MBBA} and {\it PAA} liquid crystals where ${\cal B}_{MBBA}=5.92,\;
{\cal B}_{PAA}=7.11$ (see Tables 1 and 2 in Appendix).
In the opposite case, ${\cal B}<2$, one can find the regime where the above roots are 
complex--conjugates. This gives rise to the oscillatory evolution of the jet which
appears via Hopf bifurcation (see Section \ref{cut1}).

\begin{figure}[h]
\centerline{\psfig{figure=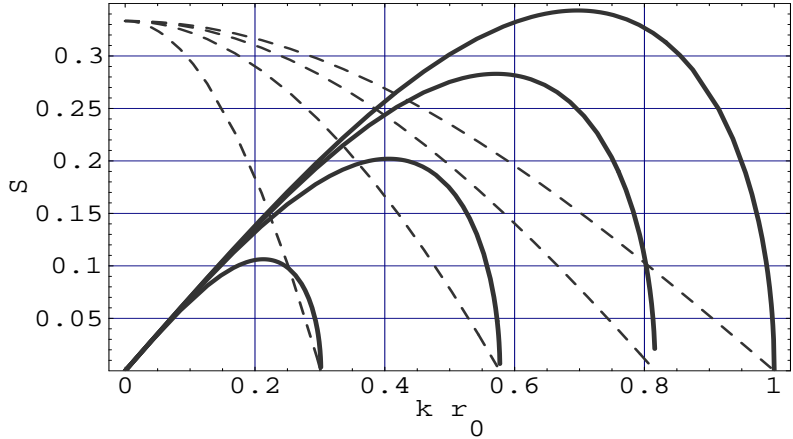,height=6cm,width=11cm}}
\vspace{-.1cm}
\caption{A plot of rescaled growth rate $S$ {\it vs.} $kr_0$ for low viscosity 
$\sqrt{\rho r_0^3/\sigma_0}\;s_{-}(kr_0)$ ({\it plain line}) and 
high viscosity $2\beta_2 r_0/\sigma_0\;s_{+}(kr_0)$ ({\it dashed line}) 
for different values of $\varkappa$ in descending order from above: 
$\varkappa=0,\;0.25,\;1,\;5$. If $\vartheta/\nu=4\varkappa$,
then the scaling for both viscous regimes is the same.}
\label{figplat2}
\end{figure}

A significant simplification can be obtained if we assume degeneration of the  
three viscosities at critical value ${\cal B}_*=2$. Indeed, when the viscous 
moduli $\beta_j$ satisfy the relation 
\begin{eqnarray}
{\cal B}_*(\beta_j)=2\;\;\longrightarrow\;\;
2\beta_1+\beta_3=3\beta_2\;,
\label{dec0}
\end{eqnarray}
the three roots $m_j^2$ of equation (\ref{disper12}) are 
\begin{eqnarray}
m_{1*}^2=k^2\;,\;\;\;m_{2*}^2=k^2+\frac{s}{\overline{\nu_2}}\;,\;\;\;m_3^2=
\frac{s}{\vartheta}\;.
\label{dec03}
\end{eqnarray}
Note that (\ref{dec0}) cancels the last term in (\ref{frank19}). The 
expressions (\ref{dec03}) indicate that the problem was decomposed in two 
parts, or, in other words, the cross--terms in equations (\ref{disper7a}), 
(\ref{disper7b}) are dropped. Thus, the first part of the problem is 
associated with Rayleigh instability as described by
\begin{eqnarray}
\left(\Delta_{2c}-m_{1*}^2\right)\left(\Delta_{2c}-m_{2*}^2\right)\chi=0\;,\;\;\;\;
\label{dec1}
\end{eqnarray}
with boundary conditions (BC) that account for the elasticity
\begin{eqnarray}
\frac{\partial }{\partial r}\Delta_{2c}\chi+
k^2\frac{\partial \chi}{\partial r}=0\;,\;\;\;\;
\frac{s}{k}\;p=2s\beta_1\frac{\partial^2 \chi}{\partial r^2}+
\Gamma\frac{\partial \chi}{\partial r}\;\;\;\mbox{at}\;\;\;r=r_0\;.
\label{dec2}
\end{eqnarray}
The second part is associated with an orientational instability of the 
director field ${\bf n}({\bf r},t)$, 
\begin{eqnarray}
\left(\Delta_{2c}-m_3^2\right)\theta=0\;,\;\;\;\;\mbox{with BC}\;\;\;\;
s\frac{\partial \theta}{\partial r}=k^2\frac{\partial \chi}{\partial r}
\;\;\;\mbox{at}\;\;\;r=r_0\;.
\label{dec3}
\end{eqnarray}
The solutions of equations (\ref{dec1}) and (\ref{dec3}) are
\begin{equation}
\chi(r)=\frac{c_1}{m_{1*}}I_0(m_{1*}r)+\frac{c_2}{m_{2*}}I_0(m_{2*}r)\;,\;\;\;\;
\theta(r)=\frac{c_3}{m_3}I_0(m_3r)\;.
\label{dec3a}
\end{equation}
Hence, using these solutions, the hydrodynamic pressure $p(r)$, stream 
function $\psi(r)$ and surface displacement $\varsigma(r_0)$ are obtained as
\begin{eqnarray}
p(r)=-c_1s\rho I_0(m_{1*}r)\;,\;\;\;\;
\psi(r)=r\left[c_1I_1(m_{1*}r)+c_2I_1(m_{2*}r)\right]\;,\;\;\;\;
\varsigma(r_0)=\frac{c_3}{k}I_1(m_3r_0)\;,\nonumber
\end{eqnarray}
where here the only indeterminate are $c_1$ and $c_2$, while $c_3$ can be expressed 
as their linear combination, 
\begin{eqnarray}
c_3\;\frac{s}{k^2}=c_1\frac{I_1(m_{1*}r_0)}{I_1(m_3r_0)}+
c_2\frac{I_1(m_{2*}r_0)}{I_1(m_3r_0)}\;,
\label{dec5}
\end{eqnarray}
provided that $s=s(kr_0)$ satisfies the dispersion relation which comes 
from (\ref{dec2}), (\ref{dec3a})
\begin{eqnarray}
s^2+\frac{2\overline{\nu_1} k^2}{I_0(kr_0)}\left[I_1'(kr_0)-
\frac{2km_{2*}}{k^2+m_{2*}^2}\frac{I_1(kr_0)}{I_1(m_{2*}r_0)}I_1'(m_{2*}r_0)\right]s=
\nonumber\\
\frac{\sigma_0k}{\rho r_0^2}
\left[1-k^2r_0^2(1+2\varkappa)\right]\frac{I_1(kr_0)}{I_0(kr_0)}
\frac{m_{2*}^2-k^2}{m_{2*}^2+k^2}\;.
\label{weber}
\end{eqnarray}
If $\varkappa=0$ and $\overline{\nu_1}=\overline{\nu_2}$, then equation (\ref{weber}) 
is known as {\it Weber equation} for a viscous isotropic liquid \cite{Levich62}. 
For low viscosity, $\beta_1\sim\beta_2\ll \sqrt{\rho \sigma_0 r_0}$, a Rayleigh type 
expression is obtained (see Figure \ref{figplat2})
\begin{eqnarray}
s^2_{-}(kr_0)=\frac{\sigma_0k}{\rho r_0^2}
\left[1-k^2r_0^2(1+2\varkappa)\right]\frac{I_1(kr_0)}{I_0(kr_0)}\;,
\label{dec6}
\end{eqnarray}
where subscript "$_{-}$" denotes low viscosity.

\noindent
The maximum $s_{-}^{max}$ in equation (\ref{dec6}) which corresponds to the 
wave number $k_{-}^{max}$, gives rise to evolution of the largest capillary 
instability. Numerical calculation shows that $s_{-}^{max}$ and 
$k_{-}^{max}$ are both proportional to $(1+2\varkappa)^{-1/2}$
\begin{eqnarray}
s_{-}^{max}\cong \frac{1}{3\sqrt{1+2\varkappa}}
\sqrt{\frac{\sigma_0}{\rho r_0^3}}\;,\;\;\;
k_{-}^{max}\cong \frac{a}{r_0\sqrt{1+2\varkappa}}\;,\;\;\;
a=0.697\;.
\label{dec6a}
\end{eqnarray}
When high viscosity prevails $\beta_1\sim\beta_2\gg \sqrt{\rho \sigma_0 r_0}$, 
the dispersion equation reads (see Figure  \ref{figplat2})
\begin{eqnarray}
s_{+}(kr_0)=\frac{\sigma_0}{2\beta_2r_0^2 k}\;
\frac{\left[1-k^2r_0^2(1+2\varkappa)\right]I_1^2(kr_0)}
{I_0(kr_0)I_1(kr_0)+kr_0\left[I_1^{'}(kr_0)\right]^2}\;,\;\;\;
s_{+}^{max}\cong \frac{\sigma_0}{6\beta_2r_0}\;,\;\;\;
k_{+}^{max}=0\;.
\label{dec7}
\end{eqnarray}
where subscript "$_{+}$" denotes high viscosity. Similar to ordinary liquids 
\cite{Chandra61}, in this limit there is no finite mode of maximum instability for 
any $\varkappa$. In this case we have
\begin{eqnarray}
\varsigma(r_0)=\frac{k_{+}^{max}}{s_{+}^{max}}
\left[c_1I_1(k_{+}^{max}r_0)+c_2I_1(m_{2*}r_0)\right]=0\;.
\label{dec7a}
\end{eqnarray}
Nevertheless, there exists a continuous range 
$[\;0,(1+2\varkappa)^{-1/2}r_0^{-1}\;]$ of wave numbers $k$, with finite 
disturbance growth rate $s_{+}(kr_0)$, which affect the cylindrical jet.

\subsection{Hydrodynamic influence on LC's orientational instability}
\label{influen}
We conclude this Section with a brief discussion regarding the hydrodynamic 
influence on the orientational instability of the director field
${\bf n}({\bf r},t)$. As the effect of hydrodynamics changes the wave number 
$k_s$ of Plateau instability to $k_{max}$, the flow drives the orientational 
instability (\ref{plat2b}) of the director field ${\bf n}({\bf r},t)$. 
Indeed, according to (\ref{dec3a})
\begin{eqnarray}
n_r^1(r,z)=c_3I_1\left(m_3^{max}r\right)\;,\;\;\;
m^{max}_3=\sqrt{\frac{s^{max}}{\vartheta}}\;.
\label{dec8}
\end{eqnarray}
It is convenient to consider the following two marginal viscous regimes. 

1. The low--viscosity limit: 
\begin{eqnarray}
\left(m_{3-}^{max}r_0\right)^2 \cong 
\frac{1}{3\sqrt{1+2\varkappa}}\frac{1}{\sqrt{\varkappa 
\varepsilon}}\;,\;\;\;\;\varepsilon=\frac{\rho K}{\gamma_1^2}\;,
\label{dec9}
\end{eqnarray}
where $\varepsilon\sim 10^{-6}\div 10^{-4}$ is a small dimensionless
parameter.

2. The high--viscosity limit: 
\begin{eqnarray}
\left(w_{3+}^{max}r_0\right)^2 \cong 
\frac{1}{6\varkappa}\frac{\gamma_1}{\beta_2}\;.
\label{dec10}
\end{eqnarray}
In both limits the distribution of director field ${\bf n}({\bf 
r},t)$ in the jet is always nontrivial and definitely far from static 
distribution (\ref{plat2b}). 

\section{Conclusion}
\label{Conc}

1. The capillary instability of liquid crystalline (LC) jet is considered
in the framework of linear hydrodynamics of uniaxial nematic LC. Its 
static version, called Plateau instability and being correspondent to the 
variational problem of minimal free energy, predicts an essential dependence of 
the disturbance {\it cut--off} wavelength upon the dimensionless parameter 
$\varkappa=K/\sigma_0 r_0$.

2. The hydrodynamic problem of capillary instability in LC jets is solved 
exactly followed by derivation of the dispersion relation. This relation, which 
is represented as a determinant equation, expresses implicitly the dispersion 
$s=s(k)$ of the growth rate $s$, as a function of the wave number $k$ of 
axisymmetric disturbances of the jet.

3. The case, where the dispersion equation becomes solvable, is considered in 
detail. It corresponds to the regime, wherein the hydrodynamic and 
orientational modes become decoupled. Hydrodynamics changes the wave number  
$k_s$ of Plateau instability into $k_{max}$ that produces evolution of the 
largest capillary instability. Similarly, hydrodynamic flow influences the static 
orientational instability of the director field ${\bf n}({\bf r},t)$.

4. The present theory can be easily extended to non--uniaxial nematic LC, which 
possesses finite point symmetry groups $G\subset O(3)$ as distinguished 
from uniaxial group ${\sf D}_{\infty h}$. The corresponding expressions for the 
free energy density $E_d(G)$ and the dissipative function $D(G)$ were derived 
in \cite{Fel89}.

5. In this work the effect of external fields was not considered. 
However, the theory developed here facilitates the treatment of Rayleigh 
instability in nematic LC in the presence of static electromagnetic 
fields.

\section{Acknowledgement}
The research was supported from Gileadi Fellowship program of the Ministry
of Absorption of the State of Israel. The useful comments of E. I. Kats 
are hereby acknowledged.

\newpage
\appendix
\renewcommand{\theequation}{\thesection\arabic{equation}}
\section{Appendix}
\label{appendix}
\setcounter{equation}{0}
\begin{center}
{\bf Table$\;$1.} The basic physical parameters $\alpha_i$, $\rho$, $K$, 
$\sigma_0$ and their derivatives $\eta_i$, $\beta_i$, $\gamma_i$, $B_i$, 
$\mu_i$, $\lambda$ and $\nu_i$ for nematic liquid crystal {\it 
4--metoxybenziliden--4--butilanilin (MBBA)} at $25^{\circ}C$ 
taken from \cite{DEJeu80}, \cite{Cogn82}.

\vspace{.2cm}
\begin{tabular}{||c|c|c|c|c|c|c|c|c|c|c||} \hline\hline
$\alpha_1,\;mPa\cdot s$ & $\alpha_2,\;mPa \cdot s$ & 
$\alpha_3,\;mPa \cdot s$ & $\alpha_4,\;mPa \cdot s$ & $\alpha_5,\;mPa 
\cdot s$ & $\alpha_6,\;mPa \cdot s$ \\ \hline
$7$ &  $-78$ & $-1$ & $84$ & $46$ & $-33$  \\ \hline\hline
$\eta_1,\;mPa\cdot s$ & $\eta_3,\;mPa\cdot s$ & $\eta_5,\;mPa\cdot s$ & 
$\lambda$ & $\mu_1$ & $\mu_2$ \\ \hline 
$42$ &$50$ & $104$ & 1.026 & 1.013 & $0.013$  \\ \hline\hline
$\beta_1,\;mPa\cdot s$ & $\beta_2,\;mPa\cdot s$ & $\beta_3,\;mPa\cdot s$ & 
$\beta_4,\;mPa\cdot s$ & $\gamma_1,\;mPa\cdot s$ & $\gamma_2,\;mPa\cdot s$ 
\\ \hline
$42$ &$25$ & $79$ & $59$ & $77$ & $-79$ \\ \hline\hline 
$B_1,\;mPa\cdot s$ & $B_2,\;mPa\cdot s$ & $B_3,\;mPa\cdot s$ & 
$B_4,\;mPa\cdot s$ & ${\cal B}$ &  $\vartheta,\;m^2/s$\\ \hline 
$58$ &  $104$ & $25$ & $78$ & $5.92$ & $1.2\times 10^{-10}$ \\ \hline\hline
$\rho,\;kg/m^3$ & $K,\;N$ & $\sigma_0,\;N/m$ & $K/\sigma_0,\;m$ & 
$\nu_i,\;m^2/s$ & $\vartheta/\nu_i$ \\ \hline
$1.2\times 10^3$ &$9\times 10^{-12}$ & $38\times 10^{-3}$ & $2.4\times 
10^{-10}$ & $10^{-5}\div 10^{-4}$ & $10^{-6}\div 10^{-5}$  \\ \hline\hline
\end{tabular}
\label{tavla1}
\end{center}

\vspace{.5cm}
\begin{center}
{\bf Table$\;$2.} The basic physical parameters $\alpha_i$, $\rho$, $K$,
$\sigma_0$ and their derivatives $\eta_i$, $\beta_i$, $\gamma_i$, $B_i$,
$\mu_i$, $\lambda$ and $\nu_i$ for nematic liquid crystal {\it 
para--azoxyanisole (PAA)} at $122^{\circ}C$ taken from 
\cite{DEJeu80}, \cite{Cogn82}.

\vspace{.2cm}
\begin{tabular}{||c|c|c|c|c|c|c|c|c|c|c||} \hline\hline
$\alpha_1,\;mPa\cdot s$ & $\alpha_2,\;mPa \cdot s$ &
$\alpha_3,\;mPa \cdot s$ & $\alpha_4,\;mPa \cdot s$ & $\alpha_5,\;mPa
\cdot s$ & $\alpha_6,\;mPa \cdot s$ \\ \hline
$4$ &  $-6.9$ & $-0.2$ & $6.8$ & $5$ & $-2.1$  \\ \hline\hline
$\eta_1,\;mPa\cdot s$ & $\eta_3,\;mPa\cdot s$ & $\eta_5,\;mPa\cdot s$ &
$\lambda$ & $\mu_1$ & $\mu_2$ \\ \hline
$3.4$ &$4.5$ & $13.7$ & 1.06 & 1.03 & $0.03$  \\ \hline\hline
$\beta_1,\;mPa\cdot s$ & $\beta_2,\;mPa\cdot s$ & $\beta_3,\;mPa\cdot s$ &
$\beta_4,\;mPa\cdot s$ & $\gamma_1,\;mPa\cdot s$ & $\gamma_2,\;mPa\cdot s$
\\ \hline
$3.4$ &$2.25$ & $11.45$ & $4.55$ & $6.7$ & $-7.1$ \\ \hline\hline
$B_1,\;mPa\cdot s$ & $B_2,\;mPa\cdot s$ & $B_3,\;mPa\cdot s$ &
$B_4,\;mPa\cdot s$ & ${\cal B}$ &  $\vartheta,\;m^2/s$\\ \hline
$4.34$ &  $9.36$ & $2.26$ & $11.24$ & $7.11$ & $1.8\times 10^{-9}$ \\ 
\hline\hline
$\rho,\;kg/m^3$ & $K,\;N$ & $\sigma_0,\;N/m$ & $K/\sigma_0,\;m$ &
$\nu_i,\;m^2/s$ & $\vartheta/\nu_i$ \\ \hline
$1.4\times 10^3$ &$11.9\times 10^{-12}$ & $40\times 10^{-3}$ & $3\times
10^{-10}$ & $10^{-6}\div 10^{-5}$ & $10^{-4}\div 10^{-3}$  \\ \hline\hline
\end{tabular}
\label{tavla2}
\end{center}



\end{document}